%% file: paper.tex
\documentclass[12pt]{elsarticle}
\usepackage{graphicx}
\usepackage{xspace}
\usepackage{amsmath}
\usepackage{amsfonts}
\usepackage{amssymb}
\usepackage{enumerate}
\usepackage{epsfig}
\usepackage{lineno}

\newcommand{\ksn}{\ensuremath{\mathrm{K_S}}\xspace}

\newcommand{\kln}{\ensuremath{\mathrm{K_L}}\xspace}

\newcommand{\kkn}{\ensuremath{\mathrm{K}}\xspace}

\newcommand{\kn}{\ensuremath{\mathrm{K^0}}\xspace}
\newcommand{\knb}{\ensuremath{\mathrm{\bar{K}^0}}\xspace}
\newcommand{\kpp}{\ensuremath{\mathrm{K_+}}\xspace}
\newcommand{\knn}{\ensuremath{\mathrm{K_-}}\xspace}
\newcommand{\kppp}{\ensuremath{\mathrm{\widetilde K_+}}\xspace}
\newcommand{\knnp}{\ensuremath{\mathrm{\widetilde K_-}}\xspace}

\newcommand{\bbn}{\ensuremath{\mathrm{B}}\xspace}

\newcommand{\kknx}{\ensuremath{\mathrm{K_X}}\xspace}
\newcommand{\kkny}{\ensuremath{\mathrm{K_Y}}\xspace}

\newcommand{\kknxb}{\ensuremath{\mathrm{\bar{K}_X}}\xspace}
\newcommand{\kknyb}{\ensuremath{\mathrm{\bar{K}_Y}}\xspace}

\include{didome_macro}
\include{babarsym}
\def\CP                {\ensuremath{\mathcal{CP}}\xspace}
\def\CPT               {\ensuremath{\mathcal{CPT}}\xspace} 

\def\T       {\ensuremath{\mathcal{T}}\xspace}
\def\K       {\ensuremath{K}\xspace}

\newcommand\trule{\rule{0pt}{2.6ex}}

\journal{Nuclear Physics B}
\begin{document}
\begin{frontmatter}
%
\title{ \flushright{\scriptsize
\small \rm FTUV-12-0803} \\ \center
Direct test of time-reversal symmetry in \\ 
the entangled neutral kaon system at a $\phi$-factory
}
\author[aa,cc]{J. Bernabeu}
\ead{jose.bernabeu@uv.es}
\author[bb]{A. Di Domenico}
\ead{antonio.didomenico@roma1.infn.it}
\author[aa]{P. Villanueva-Perez}
\ead{pablo.villanueva.perez@ific.uv.es}
\address[aa]{Department of Theoretical Physics, University of Valencia, and \\
IFIC, Univ. Valencia-CSIC, E-46100 Burjassot, Valencia, Spain}
\address[bb]{Department of Physics, Sapienza University of Rome, and \\
INFN Sezione di Roma, P.le A.~Moro, 2, I-00185 Rome, Italy}
\address[cc]{PHYS-TH, CERN, CH-1211 Geneva 23, Switzerland}

\begin{abstract}
\par
We present a novel method to perform a direct \T (time reversal) symmetry test in the neutral kaon system, 
independent of any \CP and/or \CPT symmetry tests. 
This is
based on the comparison of suitable transition probabilities, where the required interchange of {\it in} $\leftrightarrow$ {\it out} states for a given process is obtained exploiting the Einstein-Podolsky-Rosen correlations of neutral kaon pairs produced at a $\phi$-factory. 
In the time distribution between the two decays, we compare a reference transition like
the one defined by the time ordered decays ($\ell^-,\pi\pi$) with the \T-conjugated one
defined by ($3\pi^0, \ell^+$).
%
With the use of this and other \T conjugated comparisons,
the KLOE-2 experiment at DA$\Phi$NE could make
a significant test.

%
%
%
%
%

\end{abstract}


\end{frontmatter}


\section{Introduction \label{sec:introduction}}
\input intro


\section{The kaon states\label{sec:definitions}}
\input definitions

\section{Observables for the \T symmetry test\label{sec:observables}}
\input tsymmetries

\section{Measurement of $R_i $ at a $\phi$-factory\label{sec:measurement}}
\input measurement
\section{Conclusions}
\input conclusions
\section*{Acknowledgements}
This work has been supported by the MINECO and Generalitat Valenciana Grants
FPA2011-23596 and PROMETEO-2008/004.


\input biblio
\appendix
\section{Orthogonality constrains \label{sec:orthogonality}}
\input orthogonality


\end{document}

%% file: didome_macro.tex
\newcommand{\BR}[1]{\ensuremath{\mathrm{BR}(#1)}}

\newcommand{\VA}[3]{\ifthenelse{\equal{#2}{#3}}
{\ensuremath{#1\pm#2}}{\ensuremath{#1\,^{+#2}_{-#3}}}}

\def\slash#1{\setbox0 = \hbox{$#1$}#1\hskip - \wd0\hbox to\wd0{\hss\sl/\/\hss}}

\newcommand{\CP}{\ensuremath{CP}}
\newcommand{\CPT}{\ensuremath{CPT}}



%% file: intro.tex
\par

\CP violation in the Standard Model (SM) arises from the single physically relevant phase in the three families Cabibbo-Kobayashi-Maskawa (CKM) mixing matrix. The existence of this matrix conveys the fact that the quarks that participate in weak processes are a linear combination of mass eigenstates. This mechanism has been validated  in the past years of experiments probing \CP violation, especially in \kkn~\cite{ref:CPVK,ref:BannerAdlerApostolakis} and \bbn~\cite{ref:mixingInducedCP-Bs,ref:directCP-Bs} meson decays.
In the context of local quantum field theories with Lorentz invariance and Hermiticity, the \CPT theorem ensures an automatic theoretical connection between \CP symmetry and \T (time reversal)
symmetry.
Since the SM is \CPT invariant, it predicts 
\T violating effects in parallel to 
each \CP-violation effect that arises due to the interference of amplitudes with different weak phases.
\par
Even though \CPT invariance has been confirmed by all present experimental tests,
particularly in the neutral kaon system where there are strong limits to possible \CPT violation 
effects
%
\cite{ref:CPTtests,ref:pdg2010,ref:KLOE_BS,ref:KLOE06,ref:KLOE08,ref:KLOE10},
the theoretical connection between \CP and \T symmetries does not imply an experimental
identity between them, except for processes which are \CPT even, 
e.g. $\kn \rightarrow \knb$ \cite{ref:kabir}.
Therefore it is of great interest 
to search for direct evidence of non-invariance under time reversal, independent of \CP violation and \CPT invariance.
Only recently, the first direct observation of \T violation, in this sense, has been 
accomplished in the neutral $\bbn$ meson system \cite{ref:babarTviol}.
In the case of transition processes a test of 
\T non-invariance needs the comparison between the transition amplitudes under the interchange between {\it in} states and {\it out} states. 
For unstable systems, the associated irreversibility looks like it prevents a true test of \T symmetry~\cite{ref:Wolfenstein}.
\par
In this article we describe the methodology 
to perform a direct 
test of \T symmetry in the neutral $\kkn$ meson system at a 
$\phi$-factory, overcoming the irreversibility problem, similarly as described in Ref.~\cite{ref:Bmethod} for a $\bbn$-factory.
This methodology makes use of Einstein-Podolski-Rosen (EPR) entanglement~\cite{ref:EPR}, 
and
relies on the possibility of preparing the quantum mechanical individual state of the neutral \kkn meson
by the observation of particular decay channels of its orthogonal entangled partner, 
and studying the time evolution of the filtered state of the still living meson.
This strategy allows the interchange of {\it in}~$\hbox{} \leftrightarrow$~{\it out} states for a given process,
as needed for a genuine test of \T symmetry.
Whereas the basic ideas have been presented 
previously~\cite{ref:bernabeuPLB-NPB} and scrutinized later~\cite{ref:Wolfenstein,ref:QuinnDiscrete,ref:Nakada,ref:BernabeuDiscrete}, 
the discussion of the steps 
to implement these concepts into a \bbn-factory experiment able to produce the desired result
has been recently presented \cite{ref:Bmethod} and later actually observed in the neutral \bbn meson system \cite{ref:babarTviol}.
Here we discuss the corresponding concepts needed for a direct \T symmetry test in the physical context of the neutral \K meson system at a $\phi$-factory. In addition we evaluate the statistical significance of the test achievable with
 the KLOE-2 experiment at DA$\Phi$NE, the Frascati $\phi$-factory~\cite{kloe2epjc}.

%% file: definitions.tex
\par
In order to formulate a possible \T symmetry test with neutral kaons,
it is necessary to precisely define the different states involved. 
First, let us consider the physical states $|\ksn\rangle$, $|\kln\rangle$, 
i.e. the states with definite masses $m_{S,L}$ and lifetimes $\tau_{S,L}$
which evolve as a function of the kaon proper time $t$ as pure exponentials 
\begin{eqnarray}
|\ksn(t)\rangle&=&e^{-i\lambda_{S}t}|\ksn\rangle \nonumber\\
|\kln(t)\rangle&=&e^{-i\lambda_{L}t}|\kln\rangle ~.
\end{eqnarray}
with $\lambda_{S,L}=m_{S,L}-i\Gamma_{S,L}/2$, and $\Gamma_{S,L}=(\tau_{S,L})^{-1}$.
They are usually expressed in terms of the flavor 
eigenstates $|\kn\rangle$, $|\knb\rangle$ as:
\begin{eqnarray}
|\ksn\rangle &=& \frac{1}{\sqrt{2\left(1+|\epsilon_S|^2\right)}}
\left[ (1+\epsilon_S) |\kn \rangle
+(1-\epsilon_S) |\knb \rangle
\right] \\
|\kln\rangle &=& \frac{1}{\sqrt{2\left(1+|\epsilon_L|^2\right)}}
\left[ (1+\epsilon_L) |\kn \rangle
-(1-\epsilon_L) |\knb \rangle
\right]
~,
\end{eqnarray}
with $\epsilon_S$
and $\epsilon_L$
two small complex parameters describing the \CP impurity in the physical states.
One can equivalently define
${\epsilon} \equiv (\epsilon_S+\epsilon_L)/2$, and
$ \delta \equiv (\epsilon_S-\epsilon_L)/2$;
adopting a suitable phase convention 
(e.g. the Wu-Yang phase convention \cite{ref:wuyang}) 
$\epsilon\neq0$ implies \T violation,
$\delta\neq0$ implies \CPT violation,
while $\delta\neq0$ or
$\epsilon\neq0$ implies \CP violation.
\\
Let us also consider the states $|\kpp\rangle$, $|\knn\rangle$ defined as follows:
$|\kpp\rangle$ is the state filtered by the decay into $\pi\pi$ 
($\pi^+\pi^+$ or $\pi^0\pi^0$), a
pure $\CP=+1$
state; 
$|\knnp\rangle$ is the state orthogonal to $|\kpp\rangle$, i.e.
$\langle\knnp|\kpp\rangle=0$, which cannot decay into $\pi\pi$,
$\langle\pi\pi|T|\knnp\rangle=0$, and is defined by \cite{ref:lipkin}:
\begin{equation}
|\knnp\rangle \equiv {\rm\widetilde N_-} \left[| \kln\rangle 
- \eta_{\pi\pi}|\ksn \rangle \right]
\end{equation}
where $\eta_{\pi\pi}=\frac{\langle \pi\pi |T |\kln\rangle}
{\langle \pi\pi |T |\ksn\rangle}$, 
and $|\rm\widetilde N_-|^2=\left[ 1 + |\eta_{\pi\pi}|^2 
- 2\Re\left( \eta_{\pi\pi}\langle\kln |\ksn \rangle \right) \right]^{-1}$ 
defines the normalization constant up to a phase factor.
Therefore the state  $|\kpp\rangle$ can be explicitly written as the state
orthogonal  to $|\knnp\rangle$ as:
\begin{equation}
|\kpp\rangle = {\rm N_+} \left[| \ksn\rangle 
+ \alpha |\kln \rangle \right]
\end{equation}
where 
\begin{equation}
\alpha=\frac{\eta_{\pi\pi}^{\star}-\langle \kln|\ksn\rangle}
{1-\eta_{\pi\pi}^{\star}\langle \ksn|\kln\rangle}~,
\end{equation}
 and
$|\rm N_+|^2=\left[ 1 + |\alpha|^2 
+ 2\Re\left( \alpha\langle\ksn |\kln \rangle \right) \right]^{-1}$.

Analogously $|\knn\rangle$ is the state filtered by the decay into 
$3\pi^0$, a
pure $\CP=-1$
state; 
$|\kppp\rangle$ is the state orthogonal to $|\knn\rangle$, i.e.
$\langle\kppp|\knn\rangle=0$, which cannot decay into $3\pi^0$,
$\langle\pi\pi|T|\knnp\rangle=0$, and is defined by:
\begin{equation}
|\kppp\rangle \equiv {\rm\widetilde N_+} \left[| \ksn\rangle 
- \left(\eta_{3\pi^0}^{-1} \right) |\kln \rangle \right]
\end{equation}
where $\left( \eta_{3\pi^0}^{-1} \right)=\frac{\langle 3\pi^0 |T |\ksn\rangle}
{\langle 3\pi^0 |T |\kln\rangle}$, 
and $|\rm\widetilde N_+|^2=\left[ 1 + |\left(\eta_{3\pi^0}^{-1} \right)|^2 
- 2\Re\left( \left(\eta_{3\pi^0}^{-1} \right)^{\star}\langle\kln |\ksn \rangle \right) \right]^{-1}$.
Therefore the state  $|\knn\rangle$ can be explicitly written as the state
orthogonal  to $|\kppp\rangle$ as:
\begin{equation}
|\knn\rangle = {\rm N_-} \left[| \kln\rangle 
+ \beta |\ksn \rangle \right]
\end{equation}
where 
\begin{equation}
\beta=\frac{ \left( \eta_{3\pi^0}^{-1} \right)^{\star}-\langle \ksn|\kln\rangle}
{1-\left(\eta_{3\pi^0}^{-1}\right)^{\star} \langle \kln|\ksn\rangle} ~, 
\end{equation}
and
$|\rm N_-|^2=\left[ 1 + |\beta|^2 
+ 2\Re\left( \beta\langle\kln |\ksn \rangle \right) \right]^{-1}$.
\par
Even though in the following we will assume that
\begin{eqnarray}
|\kpp\rangle&\equiv&|\kppp\rangle \nonumber \\
|\knn\rangle&\equiv&|\knnp\rangle~,
\label{eq:equiv}
\end{eqnarray}
here we have kept separate definitions of 
the states $|\kpp\rangle$ and $|\knn\rangle$,
which are observed
through their decay,
from the states $|\kppp\rangle$ and $|\knnp\rangle$,
which are produced exploiting the EPR correlations in entangled
kaon pairs, 
as we will see in the next section.
\par 
Assumption (\ref{eq:equiv}) corresponds to impose
the condition of orthogonality $\langle\knn|\kpp\rangle=0$ or
$\langle\knnp|\kppp\rangle=0$. This
implies
that $\beta = -\eta_{\pi\pi}$ and $\alpha=-\left(\eta_{3\pi^0}^{-1}\right)$, which in turn
imply a precise relationship between the two amplitude ratios $\eta_{\pi\pi}$ 
and $\left(\eta_{3\pi^0}^{-1}\right)$, i.e.:
\begin{eqnarray}
\eta_{\pi\pi}&=&\frac{ \langle \ksn|\kln\rangle - \left( \eta_{3\pi^0}^{-1} \right)^{\star}}
{1-\left(\eta_{3\pi^0}^{-1}\right)^{\star} \langle \kln|\ksn\rangle} \nonumber \\
&\simeq & \langle \ksn|\kln\rangle - \left( \eta_{3\pi^0}^{-1} \right)^{\star}~,
\end{eqnarray}
or put in another form:
\begin{eqnarray}
\eta_{\pi\pi}+\left( \eta_{3\pi^0}^{-1} \right)^{\star} \simeq \langle \ksn|\kln\rangle 
\simeq
{\epsilon_L + \epsilon_S^{\star}}
\label{eq:etass}
~.
\end{eqnarray}
This equation 
clearly indicates
that we have to neglect direct \CP violation when imposing assumption
(\ref{eq:equiv}).
 In fact, for instance, eq.(\ref{eq:etass})  
cannot be simultaneously satisfied for $\pi^+\pi^+$ and $\pi^0\pi^0$
decays, being $(\eta_{\pi^+\pi^-}-\eta_{\pi^0\pi^0})=3\epsilon^{\prime}$, 
with $\epsilon^{\prime}$ the direct \CP violation parameter \cite{ref:pdg2010}.
\\
The relevance of this assumption will be discussed in
\ref{sec:orthogonality}, where it will be shown that direct \CP violation can be safely neglected
for our purposes.
%
\par
Finally we will assume the validity of the $\Delta S=\Delta Q$ rule, so that the two flavor orthogonal eigenstates $|\kn\rangle$ and $|\knb\rangle$ are identified by the charge of the lepton in semileptonic decays, i.e. a $|\kn\rangle$ can decay into $\pi^-\ell^+\nu$ and not into $\pi^+\ell^-\bar{\nu}$, and vice-versa for a $|\knb\rangle$.

%% file: tsymmetries.tex
\par
A direct evidence of \T violation would mean an experiment that, considered by itself, clearly 
shows the violation independent of and unconnected to the results of \CP violation. There 
is no existing result in the neutral $\kkn$ system that clearly demonstrates time reversal violation in this sense \cite{ref:Wolfenstein}. Sometimes the Kabir 
asymmetry $\kn \to \knb$ vs.  $\knb \to \kn$
has been presented~\cite{ref:CPLEAR,ref:CPLEAR2,ref:Nakada} as a proof for \T violation. This 
process has, however, besides the drawbacks discussed in~\cite{ref:Wolfenstein}, the feature that $\kn \to\knb$ is a \CPT even transition, so that it is impossible to separate \T violation from \CP violation 
in the Kabir asymmetry: these two transformations are experimentally identical  in
this case.

There are effects in particle physics that are odd under time $t \to -t$, but they are not 
genuine violations of time reversal \T, because do not correspond to an interchange of {\it in}-states into {\it out}-states. These kinds of $t$-asymmetries, like the macroscopic and the Universe $t$-asymmetry, can occur in theories which have an exact \T symmetry in the underlying fundamental physics~\cite{ref:QuinnDiscrete}. In fact, 
the $t$-asymmetry can only be connected~\cite{ref:bernabeuPLB-NPB} to \T asymmetry under the assumptions of \CPT 
invariance plus the absence of an absorptive part difference between the initial and final 
states of the transition. As a consequence, we have to disregard these $t$-asymmetries as 
direct evidence for \T violation. 

\par
As shown in~\cite{ref:bernabeuPLB-NPB,ref:BernabeuDiscrete}, \bbn-factories and $\phi$-factories offer the unique opportunity to show
evidence for \T violation (and \CP violation) independently from the other symmetries and to measure the corresponding effects.
The EPR entanglement  here
plays a crucial role.
Let us consider 
the neutral kaon pair produced at a $\phi$-factory in
a coherent quantum state with quantum numbers $J^{PC}=1^{- -}$ \cite{ref:HandbookAD}:
\begin{eqnarray}
  |i \rangle   =  \frac{1}{\sqrt{2}} \{ |\kn \rangle |\knb \rangle - 
 |\knb \rangle |\kn \rangle
\} 
\label{eq:state1}
\\
=  \frac{1}{\sqrt{2}} \{ |\kpp \rangle |\knn \rangle - 
 |\knn \rangle |\kpp \rangle
\label{eq:state2}
\label{eq:state3}
\}~.
\end{eqnarray}  
%
It's worth noting that one can rewrite the two particle state $|i\rangle$ in terms of 
any pair of orthogonal states of individual neutral \kkn mesons, e.g., \kn and \knb, or  \kpp and \knn defined in section~\ref{sec:definitions}.
The time evolution of
the initial state is simply given by $|i(t)\rangle=e^{-i(\lambda_S+\lambda_L)t}|i\rangle$, with $t$ common proper time of the two kaons; the initial EPR correlation given by $|i\rangle$ remains unaltered until one of the two kaons decays. 
One has also to emphasize, following what quantum mechanics dictates, that the
individual state of one neutral meson in the entangled state
is not defined before the decay process of its partner 
occurs, imposing a tag over the undecayed kaon.
Thus it is possible to have a \textquotedblleft flavor-tag\textquotedblright, i.e.
to infer the flavor (\kn or \knb) of the still alive meson by observing the specific flavor decay 
($\pi^+\ell^-\bar{\nu}$ or $\pi^-\ell^+\nu$) of the other (and first decaying) meson.
Similarly we may define a \textquotedblleft\CP-tag\textquotedblright~\cite{ref:bernabeuJHEP} as the filter imposed by the decay of one of the entangled states to a \kpp or \knn,
preparing its partner, which has not decayed yet, into the orthogonal state \knn or \kpp, respectively.
In this way we may
proceed to a partition of the complete set of events into four categories, defined by the
tag in the first decay as \kpp , \knn , \kn or \knb.
\par
Let us first consider $\kn \to \kpp$ as the reference process, by
observation of a $\pi^+ \ell^{-}\bar{\nu}$ decay at a proper time $t_1$ of the opposite \knb meson\footnote{To relax the notation we will denote
$\pi^+ \ell^- \bar{\nu}$ as $\ell^-$ and $\pi^- \ell^+ \nu$ as $\ell^+$, because of the lepton charge.}
and a $\pi\pi$ decay at a later time $t_2>t_1$, denoted as ($\ell^{-}$,$\pi\pi$), and consider:	
	\begin{enumerate}[i)]
	\item Its \T transformed $\kpp \to \kn$ $(3\pi^0,\ell^{+})$, so that the asymmetry between $\kn \to \kpp$ and $\kpp \to \kn$, as a function of $\Delta t = t_{2} - t_{1}$, is a genuine \T violating effect.
	\item Its \CP transformed $\knb \to \kpp$
 ($\ell^{+}$,$\pi\pi$), so that the asymmetry between $\kn \to \kpp$ and $\knb \to \kpp $, as a function of $\Delta t = t_{2} - t_{1}$, is a genuine \CP violating effect. 
	\item Its \CPT transformed $\kpp \to \knb$ $(3\pi^0, \ell^{-})$, so that the asymmetry between $\kn \to \kpp$and $\kpp \to \knb$, as a function of $\Delta t= t_{2} -t_{1}$, is a genuine test of \CPT invariance.
	\end{enumerate}
	


     One may check, that the events used for the asymmetries i),
       ii), and iii) are
completely independent. 

There are other three independent comparisons between \T-conjugated processes, as summarized in Table~\ref{tab:Tprocesses}. Analogously, we can apply the same methodology for similar tests of \CP violation and \CPT invariance. Tables~\ref{tab:CPprocesses} and~\ref{tab:CPTprocesses} summarize all the possible comparisons of \CP- and \CPT-conjugated transitions with their corresponding decay products.

\begin{table}[h]
  \begin{center}
    \begin{tabular}{cc|cc}		
      \hline
      \multicolumn{2}{c|}{Reference}  &   \multicolumn{2}{c}{\T-conjugate} \\
      Transition & Decay products        &   Transition &  Decay products  \\ \hline
      \trule $\kn \to \kpp$   &  $ (\ell^-, \pi\pi)$ & $\kpp \to \kn$    & $(3\pi^0, \ell^+)$ \\
      \trule $\kn \to \knn$ &  $  (\ell^-,3\pi^0)$  &  $\knn \to \kn$ & $(\pi\pi,\ell^+)$ \\
      \trule $\knb \to \kpp$   &  $ (\ell^+,\pi\pi)$ & $\kpp \to \knb$    & $(3\pi^0,\ell^-)$ \\
      \trule $\knb \to \knn$  &  $ (\ell^+,3\pi^0)$  & $\knn \to \knb$   & $(\pi\pi,\ell^-)$ \\ \hline
   \end{tabular}
    \caption{Possible comparisons between \T-conjugated transitions and the associated time-ordered decay products in the experimental $\phi$-factory scheme.
      \label{tab:Tprocesses}}
  \end{center}	
\end{table}

\begin{table}[h]
  \begin{center}
    \begin{tabular}{cc|cc}		
      \hline
      \multicolumn{2}{c|}{Reference}  &   \multicolumn{2}{c}{\CP-conjugate} \\
      Transition & Decay products       &   Transition &  Decay products  \\ \hline
      \trule $\kn \to \kpp$   &  $ (\ell^-, \pi\pi)$ & $\knb \to \kpp$    & $(\ell^+,\pi\pi)$ \\
      \trule $\kn \to \knn$ &  $  (\ell^-,3\pi^0)$  &  $\knb \to \knn$ & $(\ell^+,3\pi^0)$ \\
      \trule $\knb \to \kpp$   &  $ (\ell^+,\pi\pi)$ & $\kn \to \kpp$    & $(\ell^-,\pi\pi)$ \\
      \trule $\knb \to \knn$  &  $ (\ell^+,3\pi^0)$  & $\kn \to \knn$   & $(\ell^-,3\pi^0)$ \\ \hline
   \end{tabular}
    \caption{Possible comparisons between \CP-conjugated transitions and the associated time-ordered decay products in the experimental $\phi$-factory scheme.
      \label{tab:CPprocesses}}
  \end{center}	
\end{table}

\begin{table}[h]
  \begin{center}
    \begin{tabular}{cc|cc}		
      \hline
      \multicolumn{2}{c|}{Reference}  &   \multicolumn{2}{c}{\CPT-conjugate} \\
      Transition & Decay products        &   Transition &  Decay products  \\ \hline
      \trule $\kn \to \kpp$   &  $ (\ell^-, \pi\pi)$ & $\kpp \to \knb$    & $(3\pi^0, \ell^-)$ \\
      \trule $\kn \to \knn$ &  $  (\ell^-,3\pi^0)$  &  $\knn \to \knb$ & $(\pi\pi,\ell^-)$ \\
      \trule $\knb \to \kpp$   &  $ (\ell^+,\pi\pi)$ & $\kpp \to \kn$    & $(3\pi^0,\ell^+)$ \\
      \trule $\knb \to \knn$  &  $ (\ell^+,3\pi^0)$  & $\knn \to \kn$   & $(\pi\pi,\ell^+)$ \\ \hline
   \end{tabular}
    \caption{Possible comparisons between \CPT-conjugated transitions and the associated time-ordered decay products in the experimental $\phi$-factory scheme.
      \label{tab:CPTprocesses}}
  \end{center}	
\end{table}

\par
Our goal is to demonstrate and measure the violation of time reversal invariance.
Therefore 
we have to consider
the following ratios of probabilities:
\begin{eqnarray}
R_1(\Delta t) &=& P\left[\kn(0)\to\kpp(\Delta t)\right]/P\left[\kpp(0)\to\kn(\Delta t)\right] \nonumber \\
R_2(\Delta t) &=& P\left[\kn(0)\to\knn(\Delta t)\right]/P\left[\knn(0)\to\kn(\Delta t)\right] \nonumber\\
R_3(\Delta t) &=& P\left[\knb(0)\to\kpp(\Delta t)\right]/P\left[\kpp(0)\to\knb(\Delta t)\right] \nonumber\\
R_4(\Delta t) &=& P\left[\knb(0)\to\knn(\Delta t)\right]/P\left[\knn(0)\to\knb(\Delta t)\right]~.
\label{eq:ratios}
\end{eqnarray}
The measurement of any deviation from the prediction 
\begin{eqnarray}
R_1(\Delta t)=R_2(\Delta t)=R_3(\Delta t)=R_4(\Delta t)=1
\label{eq:tprediction}
\end{eqnarray}
imposed by \T invariance
is a signal of \T violation. This outcome will be highly
rewarding as a model-independent and a direct observation of \T violation. 
\par
If we express two generic orthogonal basis $\{ \kknx , \kknxb \}$ and $\{ \kkny , \kknyb \}$,
which in our case correspond to $\{ \kn , \knb \}$ or $\{ \kpp , \knn \}$, as follows:
%
\begin{eqnarray}
|\kknx\rangle &=& X_S| \ksn\rangle + X_L |\kln \rangle \\
|\kknxb\rangle &=& \bar{X}_S| \ksn\rangle + \bar{X}_L |\kln \rangle \\
\nonumber \\
|\kkny\rangle &=& Y_S| \ksn\rangle + Y_L |\kln \rangle \\
|\kknyb\rangle &=& \bar{Y}_S| \ksn\rangle + \bar{Y}_L |\kln \rangle ~.
\end{eqnarray}

the generic quantum mechanical expression for the probabilities entering in eqs.(\ref{eq:ratios}) is given by
\begin{eqnarray}
\label{eq:probability}
P\left[\kknx (0) \to\kkny (\Delta t)\right]  & = & \left| \langle \kkny | \kknx(\Delta t) \rangle \right|^2 \\
& = &
\frac{1}{|\det Y|^2} 
\left|
e^{-i\lambda_S \Delta t} X_S \bar{Y}_L -e^{-i\lambda_L \Delta t} X_L \bar{Y_S}
\right|^2
\nonumber \\
& = & \frac{1}{|\det Y|^2} \{e^{-\Gamma_S \Delta t} |X_S \bar{Y}_L|^2 + e^{-\Gamma_L \Delta t} |X_L \bar{Y}_S|^2 \nonumber \\
& & - 2 e^{-\frac{(\Gamma_S+\Gamma_L)}{2}\Delta t}\Re \left( e^{i\Delta m \Delta t}X_S \bar{Y}_L X_L^{\star} \bar{Y}_S^{\star} \right) \}\nonumber~,
\end{eqnarray}
with
\begin{eqnarray}
\det Y & = & Y_S \bar{Y}_L - Y_L\bar{Y}_S  
\end{eqnarray}
and
\begin{eqnarray}
\left| \det Y \right|^2  = \left| \det X \right|^2 =  \frac{1}{1-|\langle \ksn|\kln \rangle|^2}~.
\end{eqnarray}
Its inverse $P\left[\kkny (0) \to\kknx (\Delta t)\right]$ is obtained
simply with the substitution $X \leftrightarrow Y$.

%
%

\par

\begin{figure}[htbp] 
   \centering
   \includegraphics[width=5.5in]{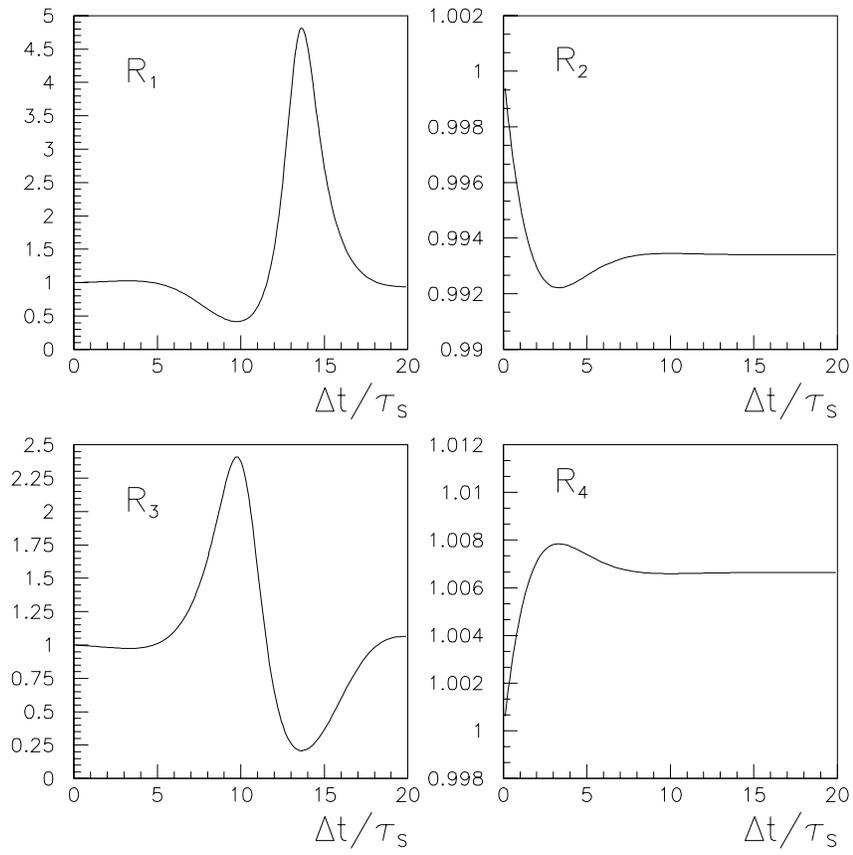} 
   \caption{The ratios $R_i$ as a function of $\Delta t$;
$R_1$ top left,
$R_2$ top right,
$R_3$ bottom left,
$R_4$ bottom right~.}
   \label{fig:fig1}
\end{figure}

Using the expected values for the $X_{S,L}$, $\bar{X}_{S,L}$, $Y_{S,L}$ and $\bar{Y}_{S,L}$ coefficients in terms of the measured 
$\epsilon$ and $\delta$ parameters \cite{ref:pdg2010},
it can be easily demostrated that the ratios $R_i$ depend on $\Delta t$, 
as it is shown in Fig.\ref{fig:fig1}.
This result is in contrast with the Kabir \T-violating asymmetry \cite{ref:kabir,ref:CPLEAR,ref:CPLEAR2}, which is independent of time:
\begin{eqnarray}
\frac{P\left[\kn (0) \to\knb (\Delta t)\right]}{P\left[\knb (0) \to\kn (\Delta t)\right]}  
 & =& \frac{
\left|X_S X_L e^{-i\lambda_S \Delta t}  -X_L X_S e^{-i\lambda_L \Delta t}\right|^2
}{
\left| \bar{X}_S \bar{X}_L e^{-i\lambda_S \Delta t}  -\bar{X}_L \bar{X}_S e^{-i\lambda_L \Delta t}\right|^2
}
=
\frac{
\left|X_S X_L \right|^2
}{
\left| \bar{X}_S \bar{X}_L \right|^2
}
\nonumber\\
&\simeq& 
\frac{
(1-4\Re\epsilon) }{(1+4\Re\epsilon)}\simeq 1-8\Re\epsilon
~.
\end{eqnarray}
\\
It is worth noting that for $\Delta t=0$ we have:
\begin{eqnarray}
R_1(0)=R_2(0)=R_3(0)=R_4(0)=1
\label{eq:one}
\end{eqnarray}
and for $\Delta t\gg \tau_S$ :
\begin{eqnarray}
R_2(\Delta t \gg \tau_S) \simeq \frac{1-2 \Re \epsilon_S}{1+2\Re\epsilon_L} \simeq 1-4 \Re \epsilon
\label{eq:tendtoconst1}
\\
R_4(\Delta t \gg \tau_S) \simeq \frac{1+2 \Re \epsilon_S}{1-2\Re\epsilon_L} \simeq 1+4\Re \epsilon
\label{eq:tendtoconst2}
\end{eqnarray}

%% file: measurement.tex
\par
\par
From the experimental point of view the observable quantity at a $\phi$-factory is the
double differential decay rate of the state $|i\rangle$
into decay products
$f_1$ 
and $f_2$ 
at proper times $t_1$ and $t_2$, respectively \cite{ref:HandbookAD}. For the time evolution of the system it is convenient to rewrite the entangled state $|i\rangle$ as:
\begin{eqnarray}
| i \rangle = \frac{\mathcal{N}}{\sqrt{2}} \{ |\ksn \rangle |\kln \rangle - 
 |\kln \rangle |\ksn \rangle \}
\label{eq:state44}
\end{eqnarray}
with
$|\mathcal{N}|^2 = \left| \det X \right|^2 
={\left[(1+|\epsilon_S|^2)(1+|\epsilon_L|^2)\right]}  / {(1-\epsilon_S\epsilon_L)^2} 
\simeq 1$
a normalization factor.
The double differential decay rate 
is given by:
\begin{eqnarray}
\label{eq:intensity}
  I(f_1,t_1;f_2,t_2)   
=  C_{12} \{ |\eta_1|^2 e^{-\Gamma_L t_1 -\Gamma_S t_2}
+|\eta_2|^2 e^{-\Gamma_S t_1 -\Gamma_L t_2} \nonumber \\
-2 |\eta_1||\eta_2|e^{-{{(\Gamma_S+\Gamma_L)}\over{2}}(t_1+t_2)}\cos[ \Delta m 
(t_1-t_2) +\phi_2-\phi_1]
\}
\end{eqnarray}  
where 
\begin{eqnarray}
\label{eq:etas}
\eta_i \equiv |\eta_i|e^{i\phi_i} = 
\frac{\langle f_i |T |\kln\rangle}{\langle f_i |T |\ksn\rangle}~,
\end{eqnarray}  
$$ C_{12}={|\mathcal{N}|^2\over 2}| \langle f_1 |T|\ksn \rangle \langle f_2 |T|\ksn
\rangle |^2~.$$

After integration on $t_1$ at fixed time difference $\Delta t=t_2-t_1>0 $, 
the decay intensity (\ref{eq:intensity}) can be rewritten 
in a more suitable form, putting in evidence the probabilities we are aiming for.
 In particular it will be a function of the first 
decay product
 $f_1=f_{\bar{X}}$ (which takes place at time $t_1$, identifies a $\kknxb$ state, and tags a $\kknx$ state on the opposite side), the second 
decay products
 $f_2=f_{Y}$ (which takes place at time $t_2$ and identifies a $\kkny$ state):
\begin{eqnarray}
I(f_{\bar{X}},f_{Y};\Delta t)&=& \int^{\infty}_0 I(f_{\bar{X}}, t_1;f_{Y};t_2) d t_1
\nonumber \\
&=&\frac{1}{\Gamma_S+\Gamma_L} \left|  
\langle \kknx \kknxb | i \rangle \langle f_{\bar{X}} |T | \kknxb  \rangle
\langle \kkny | \kknx(\Delta t) \rangle
\langle f_Y | T | \kkny  \rangle
\right|^2 \nonumber \\
&=&C(f_{\bar{X}},f_Y)\times P\left[\kknx(0) \to \kkny(\Delta t) \right]~,
\end{eqnarray}
where the coefficient $C(f_{\bar{X}},f_Y)$, depending only on the final states $f_{\bar{X}}$ and $f_Y$, is 
given by:
\begin{eqnarray}
C(f_{\bar{X}},f_Y)&=&\frac{1}{2(\Gamma_S+\Gamma_L)} \left|  
\langle f_{\bar{X}} | T |  \kknxb  \rangle
\langle f_Y | T |\kkny  \rangle
\right|^2 \nonumber \\
&=&\dfrac{\left|\langle f_{\bar{X}}|T|\ksn\rangle\right|^2\left|\langle f_{Y}|T|\ksn\rangle\right|^2}{2(\Gamma_S+\Gamma_L)}
\times \left|(\bar{X}_S+\eta_{\bar{X}}\bar{X}_L)(Y_S+\eta_{Y} Y_L)\right|^2~, \nonumber\\
\label{eq:coeffi}
\end{eqnarray}
and the generic probability $P\left[\kknx(0) \to \kkny(\Delta t) \right]$, containing the only time dependence, is the one given by eq.(\ref{eq:probability}).
%
%
%
%
\\
From eq.(\ref{eq:coeffi}) and
the condition:
\begin{eqnarray}
I(f_{\bar{X}},f_{Y};\Delta t=0)&=& C(f_{\bar{X}},f_Y)\times \left| \langle \kkny | \kknx \rangle \right|^2
\nonumber \\
&=&C(f_Y,f_{\bar{X}})\times \left| \langle \kknxb | \kknyb \rangle \right|^2
~,
\end{eqnarray}
it can be easily shown that the coefficient $C(f_{\bar{X}},f_Y)$ is invariant under interchange
$f_Y \leftrightarrow f_{\bar{X}}$, i.e. 
\begin{eqnarray}
\label{eq:coeffexchange}
C(f_{\bar{X}},f_Y)=C(f_Y,f_{\bar{X}})~.
\end{eqnarray}

One can define the following observable ratios:
\begin{eqnarray}
\label{eq:intensity2}
R_1^{\rm{exp}}(\Delta t) \equiv
\frac{  I(\ell^-,\pi\pi;\Delta t)}
{ I(3\pi^0,\ell^+;\Delta t)}   
=R_1(\Delta t) \times \frac{C(\ell^-,\pi\pi)}{C(3\pi^0,\ell^+)}
\end{eqnarray}  
\begin{eqnarray}
\label{eq:intensity2}
R_2^{\rm{exp}}(\Delta t) \equiv
\frac{  I(\ell^-,3\pi^0;\Delta t)}
{ I(\pi\pi,\ell^+;\Delta t)}   
=R_2(\Delta t)\times \frac{C(\ell^-,3\pi^0)}{C(\pi\pi,\ell^+)}
\end{eqnarray}  

\begin{eqnarray}
\label{eq:intensity2}
R_3^{\rm{exp}}(\Delta t) \equiv
\frac{  I(\ell^+,\pi\pi;\Delta t)}
{ I(3\pi^0,\ell^-;\Delta t)}   
= R_3(\Delta t)\times \frac{C(\ell^+,\pi\pi)}{C(3\pi^0,\ell^-)}
\end{eqnarray}  
\begin{eqnarray}
\label{eq:intensity2}
R_4^{\rm{exp}}(\Delta t) \equiv
\frac{  I(\ell^+,3\pi^0;\Delta t)}
{ I(\pi\pi,\ell^-;\Delta t)}   
= R_4(\Delta t)\times \frac{C(\ell^+,3\pi^0)}{C(\pi\pi,\ell^-)}~,
\end{eqnarray}  
which are proportional to the corresponding $R_i(\Delta t)$ ratios.
\par
\begin{figure}[htbp] 
   \centering
   \includegraphics[width=5.5in]{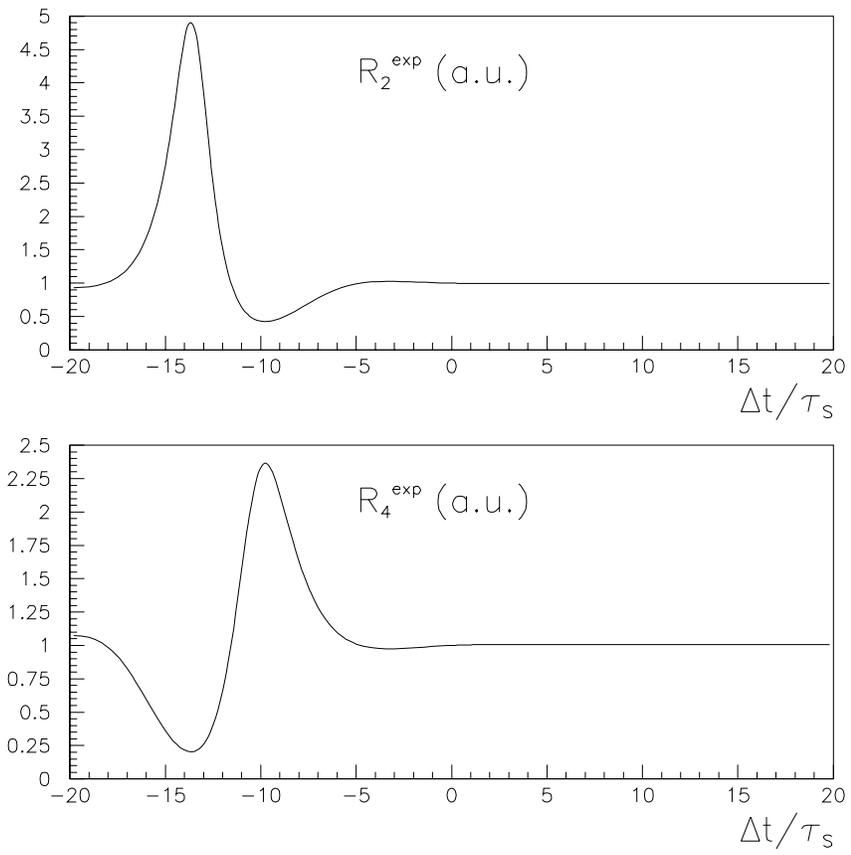} 
   \caption{The ratios $R_2^{\rm{exp}} $ and 
$R_4^{\rm{exp}}$
as a function of $\Delta t$.}
   \label{fig:fig2}
\end{figure}

It should be noted that when we perform a measurement with 
decay products in inverse time order $(f_2, f_1)$ or
$t_1 > t_2$, i.e. $\Delta t \to -\Delta t$, we are
actually measuring the inverse of {\it another} ratio, i.e.:
\begin{eqnarray}
R_2^{\rm{exp}}(-\Delta t) &=& \frac{1}{R_3^{\rm{exp}}(\Delta t)} = 
\frac{1}{R_3(\Delta t)}\times \frac{C(3\pi^0,\ell^-)}{C(\ell^+,\pi\pi)}
\\
R_4^{\rm{exp}}(-\Delta t) &=& \frac{1}{R_1^{\rm{exp}}(\Delta t)} = 
\frac{1}{R_1(\Delta t)} \times \frac{C(3\pi^0,\ell^+)}{C(\ell^-,\pi\pi)}
~.
\end{eqnarray}
Due to the property (\ref{eq:coeffexchange}), the proportionality constant
between $R_{2(4)}^{\rm{exp}}(-\Delta t) $ and $1/ R_{3(1)}(\Delta t) $
is the same as the one between  $R_{2(4)}^{\rm{exp}}(\Delta t) $ and $R_{2(4)}(\Delta t) $.
Therefore one can actually measure only two observables,
$R_2^{\rm{exp}}(\Delta t) $ and \\ 
$R_4^{\rm{exp}}(\Delta t) $, with 
$-\infty <\Delta t <+\infty$; their expected behavior is shown in Fig.\ref{fig:fig2}.
%

\par
%
%
%
From the point of view of a model independent and direct test of \T symmetry, 
it would be sufficient to prove that one of the predictions in eq.(\ref{eq:tprediction}) 
is not satisfied, i.e. that
$R_i(\Delta t)\neq 1$, for any ratio $R_i$.
Experimentally one can adopt two different strategies to obtain this result:
\begin{enumerate}
\item 
The first one is to observe any significant dependence on $\Delta t$
in the measured ratio $R_2^{\rm{exp}}(\Delta t)$ or $R_4^{\rm{exp}}(\Delta t)$; therefore one may conclude that the corresponding ratio $R_i$ is not constant and cannot satisfy the prediction in eq.(\ref{eq:tprediction}).
\item 
The second strategy consists in measuring
the ratio $R_2^{\rm{exp}}(\Delta t)$ or $R_4^{\rm{exp}}(\Delta t)$ in the limit 
$\Delta t \gg \tau_S$, where they are expected to have a constant value; given an independent evaluation of the corresponding ratio of coeffiecients 
$\frac{C(\ell^-,3\pi^0)}{C(\pi\pi,\ell^+)}$ or  $\frac{C(\ell^+,3\pi^0)}{C(\pi\pi,\ell^-)}$
one may extract the asymptotic value $R_2(\Delta t \gg \tau_S)$ or $R_4(\Delta t \gg \tau_S)$ and verify the predicted deviation from one, eq.(\ref{eq:tendtoconst1}) or (\ref{eq:tendtoconst2}).
\end{enumerate}
For the second strategy we can consider that:
\begin{eqnarray}
\frac{C(\ell^-,3\pi^0)}{C(\pi\pi,\ell^+)}&=&\left| 
\frac{ \langle \ell^-|T| \knb \rangle \langle 3\pi^0 | T | \knn \rangle }{\langle \ell^+| T | \kn \rangle \langle \pi\pi | T | \kpp \rangle}  \right|^2
\nonumber \\
&=&
\left| 
\frac{ \langle 3\pi^0 | T |  \knn \rangle }{ \langle \pi\pi | T |  \kpp \rangle}  \right|^2
\end{eqnarray}
\begin{eqnarray}
\frac{C(\ell^+,3\pi^0)}{C(\pi\pi,\ell^-)}&=&\left| 
\frac{ \langle \ell^+ | T |  \kn \rangle \langle 3\pi^0 | T | \knn \rangle }{\langle \ell^- | T |\knb \rangle \langle \pi\pi | T |  \kpp \rangle}  \right|^2
\nonumber \\
&=&
\left| 
\frac{ \langle 3\pi^0 | T | \knn \rangle }{ \langle \pi\pi | T | \kpp \rangle}  \right|^2
\label{eq:bramp2}
\end{eqnarray}
neglecting possible \CPT violation effects in semileptonic decays.
%
\par
Neglecting second order terms, one has:
\begin{eqnarray}
\label{eq:bramp1}
{\rm BR}\left( \ksn\rightarrow \pi\pi\right) \Gamma_S &=&\left| \langle \pi\pi | T | \ksn \rangle \right|^2
=\left| \frac{\langle \pi\pi |T | \kpp \rangle}{\rm N_+(1-\alpha\beta)}\right|^2 \nonumber\\
&\simeq& \left| \langle \pi\pi |T| \kpp \rangle \right|^2\\
{\rm BR}\left( \kln\rightarrow 3\pi^0\right) \Gamma_L &=&\left| \langle 3\pi^0 |T| \kln \rangle \right|^2
= \left|  \frac{\langle 3\pi^0 |T| \knn \rangle}{\rm N_-(1-\alpha\beta)}\right|^2 \nonumber \\
& \simeq & \left| \langle 3\pi^0 |T| \knn \rangle \right|^2~.
\label{eq:bramp2}
\end{eqnarray}
Using the above relations,
eqs.(\ref{eq:bramp1}) and (\ref{eq:bramp2}), one has:
\begin{eqnarray}
\frac{C(\ell^-,3\pi^0)}{C(\pi\pi,\ell^+)}\simeq \frac{C(\ell^+,3\pi^0)}{C(\pi\pi,\ell^-)}
\simeq
\frac{{\rm BR}\left( \kln\rightarrow 3\pi^0\right)  }{{\rm BR}\left( \ksn\rightarrow \pi\pi\right)  }\frac{\Gamma_L}{\Gamma_S}~.
\label{eq:brs}
\end{eqnarray}
Therefore in the case of the second strategy,
one can evaluate the ratio of coefficients in terms
of measurable branching ratios, and convert with the correct normalization the measured ratios $R_2^{\rm exp}$ and $R_4^{\rm exp}$ into the corresponding values for  $R_2$ and 
$R_4$, making possible a direct comparison of these values with 
the prediction (\ref{eq:tprediction})  obtained in the case of 
\T symmetry invariance.
\par
\begin{figure}[htbp] 
   \centering
   \includegraphics[width=5.4in]{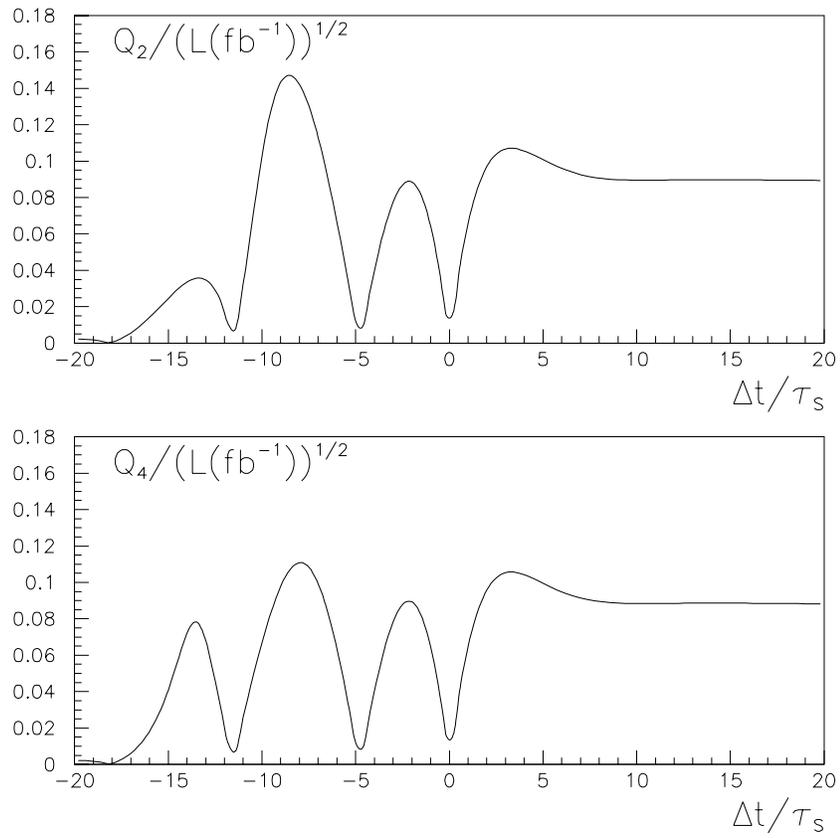} 
   \caption{
 The statistical sensitivity $Q_2(\Delta t)$ (top) and $Q_4(\Delta t)$ (bottom)
 as a function of $\Delta t$
 and normalized to the square root of the integrated luminosity $\sqrt{L{\rm (fb^{-1})}}$.
}
   \label{fig:fig5}
\end{figure}
\begin{figure}[htbp] 
   \centering
   \includegraphics[width=5.4in]{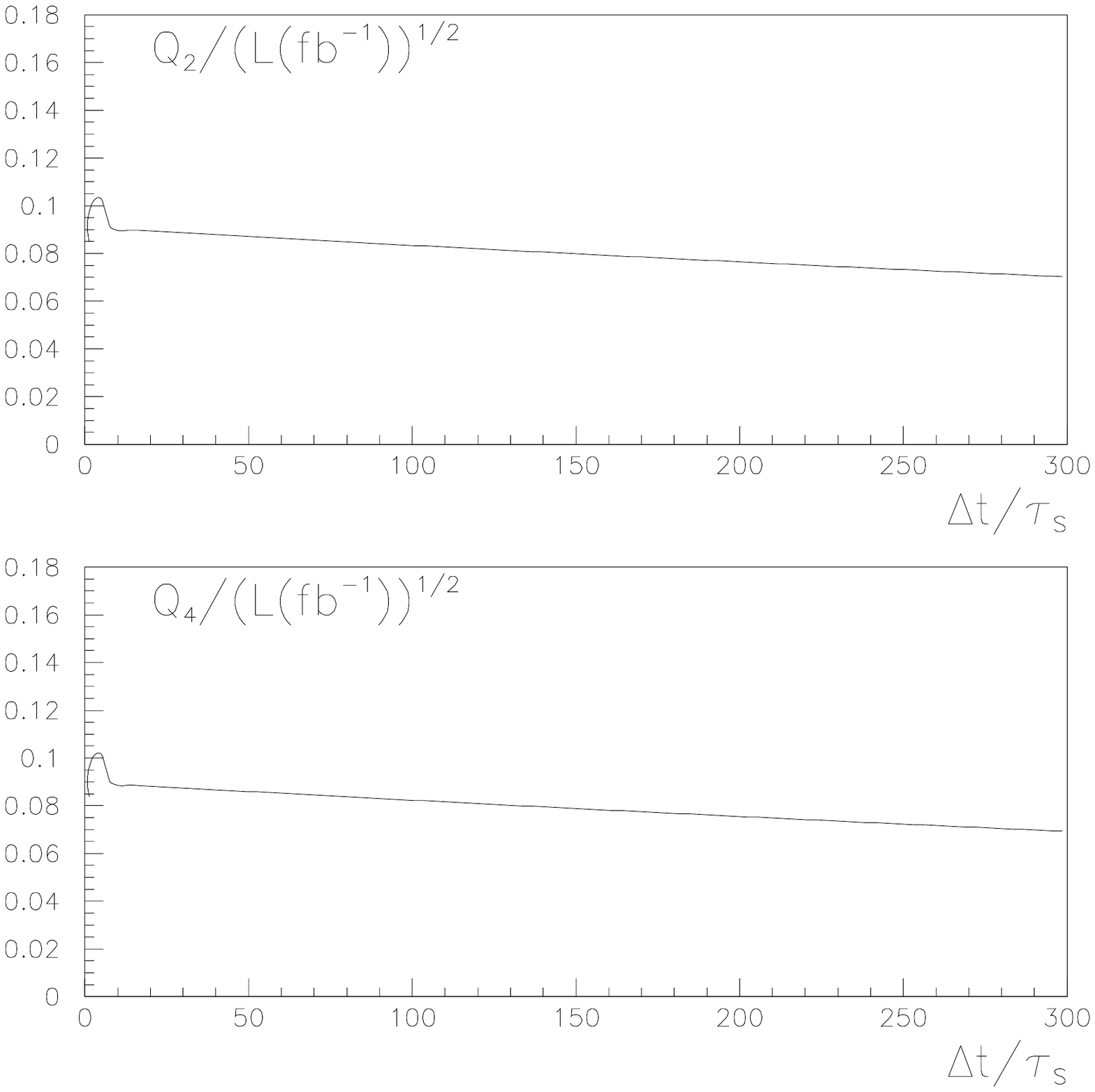} 
   \caption{As in Fig.\ref{fig:fig5} but in the range $0<\Delta t <300$ $\tau_S$.
}
   \label{fig:fig6}
\end{figure}

One can define the statistical sensitivity of an experiment
\begin{eqnarray}
Q_i(\Delta t) 
\equiv 
\frac{\left|1-R_i(\Delta t) \right|}{\sigma \left( R_i(\Delta t) \right)  }
~,
\end{eqnarray}
as the ratio between the expected deviation of $R_i$ from prediction (\ref{eq:tprediction}), as given by the measured value of $\epsilon$, and the statistical uncertainty 
on $R_i$, 
in a bin width of 1~$\tau_S$ centered at the value $\Delta t$, 
as shown in Figs.~\ref{fig:fig5} and \ref{fig:fig6}
\footnote{The plots in  Figs.~\ref{fig:fig5} and \ref{fig:fig6} have been evaluated
assuming a large number of counts and Poisson fluctuations in each $\Delta t$ bin of the measured  $I(f_1,f_2;\Delta t)$ distributions, 
and negligible uncertainties due to the knowledge of the ratio of coefficients (\ref{eq:brs})
(needed for the second strategy).
}
. 
%
%
%

It is worth noting that the sensitivity of the test in the region $\Delta t < 0$ is limited by the
large statistical uncertainties on $R_i$ (due to a fast exponential decrease of the events
in this region)
despite the expected large deviations of $R_i$ from unity.
On the other hand, in the statistically most populated region at $\Delta t >0$, the
sensitivity is not large
because small deviations of $R_i$ from unity are  expected here
(see eqs.(\ref{eq:tendtoconst1}) and (\ref{eq:tendtoconst2})).

\par
\begin{figure}[htbp] 
   \centering
   \includegraphics[width=5.4in]{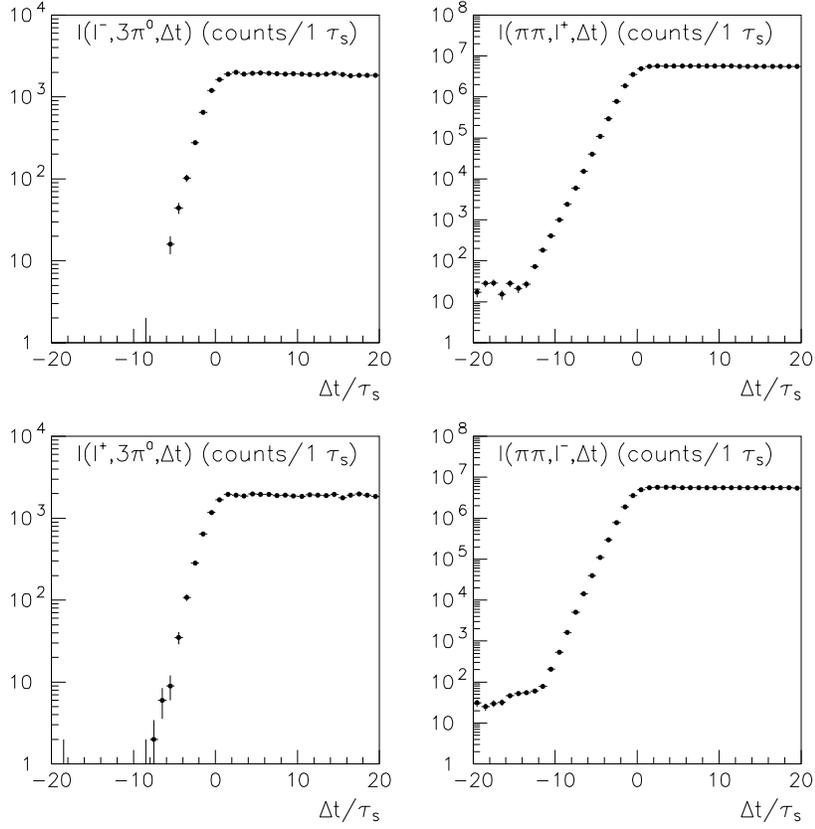} 
   \caption{The
 $I(\ell^-,3\pi^0; \Delta t)$ (top left),  
 $I(\pi\pi,\ell^+; \Delta t)$ (top right),  
 $I(\ell^+,3\pi^0; \Delta t)$ (bottom left),  and
 $I(\pi\pi,\ell^-; \Delta t)$ (bottom right) 
 distributions as a function of $\Delta t$
 evaluated with a simple Monte Carlo simulation,
making the approximation of a gaussian $\Delta t$ experimental resolution
with $\sigma=1~\tau_S$, a full detection efficiency, and 
assuming $L=10\hbox{ fb}^{-1} $. 
}
   \label{fig:fig7}
\end{figure}

In the case of the KLOE-2 experiment at DA$\Phi$NE, where an integrated luminosity $L$ of
$\mathcal{O}(10 \hbox{ fb}^{-1}  )$ is expected~\cite{kloe2epjc},
the $I(f_1,f_2 ; \Delta t)$ distributions 
have been evaluated with a simple Monte Carlo simulation,
making the approximation of a gaussian $\Delta t$ experimental resolution
with $\sigma=1~\tau_S$, and a full detection efficiency, 
as shown
in Fig.~\ref{fig:fig7}.
It can be noticed that
the $I(\ell^{\pm}, 3\pi^0; \Delta t)$ distributions
have very few or no events  for $\Delta t \lesssim-5~\tau_S$.
While a complete feasibility study is beyond the scope of the present paper, 
it appears that
the first strategy described above is difficult to be implemented 
at KLOE-2 due to the lack of enough statistics,
whereas the second 
strategy  
is much more viable.
In fact considering a large 
$\Delta t$ interval in the statistically most populated region,
e.g. 
$0\leq \Delta t \leq 300~\tau_S$,
a much larger global sensitivity of $Q \simeq 4.4, 6.2$, and $8.8$ is obtained for $L=$~5, 10, and 20~$\hbox{ fb}^{-1} $, respectively.

%
%
%
%

%% file: conclusions.tex
\par
It has been shown that, by exploiting the EPR entanglement of neutral kaon pairs
produced at a $\phi$-factory, it is possible to perform a direct test of the time reversal symmetry in the neutral kaon system, independently from \CP violation and  \CPT invariance constraints, and therefore overcoming some conceptual difficulties affecting previous tests.
The proposed test is highly model-independent, relying only on the validity
of quantum mechanical prescriptions and EPR correlations.
%
From the experimental point of view, the test would require to measure
ratios of intensities (\ref{eq:intensity}) with a suitable choice of 
decay products in definite time ordering.
The absolute normalization of the measured ratios 
requires the knowledge of measurable kaon branching ratios and lifetimes
and would not suffer from other uncertainties.
The KLOE-2 experiment at the DA$\Phi$NE 
$\phi$-factory could make a significant \T symmetry test 
with an
integrated luminosity of
$\mathcal{O}(10 \hbox{ fb}^{-1}  )$.

%% file: orthogonality.tex
\par
The orthogonality assumption (\ref{eq:equiv}) and condition (\ref{eq:etass}) constrain the $\eta_{\pi\pi}$ and
$\left(\eta_{3\pi^0}^{-1}\right)$ parameters. Concerning the  $\eta_{\pi\pi}$ parameter
one could safely neglect any contribution from direct \CP violation, because 
$(\epsilon^{\prime}/\epsilon)$ is experimentally known to be $\mathcal{O}(10^{-3})$ \cite{ref:pdg2010}. 
One can also safely neglect possible contributions
from direct \CPT violation in the $\pi\pi$ decay.
Therefore for the purposes of the present test, one can assume 
$\eta_{\pi\pi}\simeq\epsilon_L$  (e.g. adopting the Wu-Yang phase convention).
\par
\begin{figure}[htbp] 
   \centering
   \includegraphics[width=5.4in]{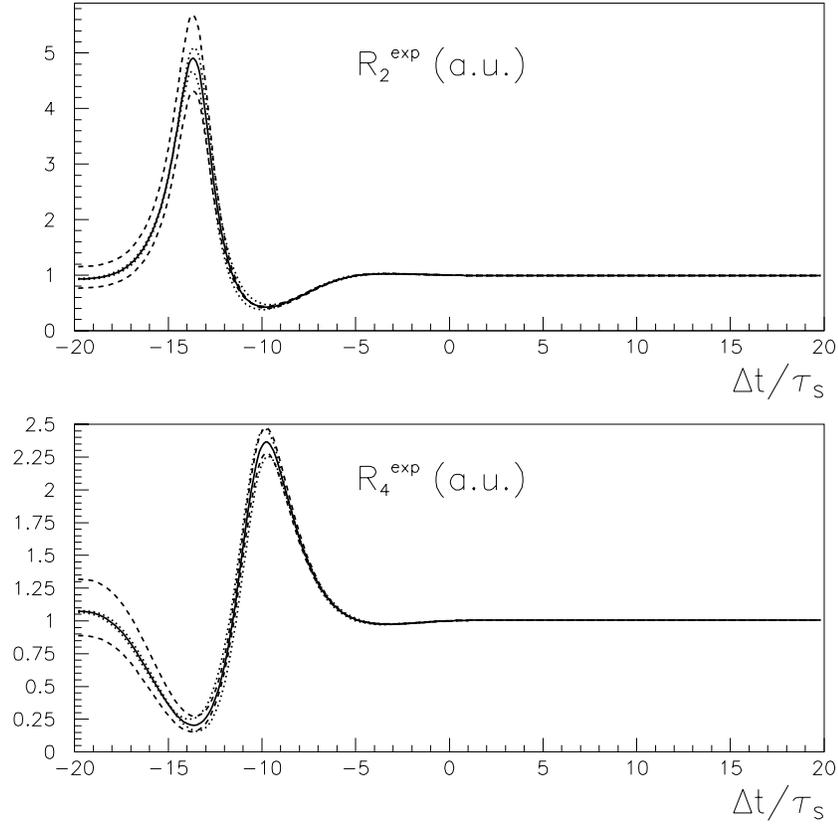} 
   \caption{The expected ratios $R_2^{\rm{exp}}(\Delta t)$ (top) and $R_4^{\rm{exp}}(\Delta t)$ (bottom) 
   as a function of $\Delta t$ (solid line); dashed lines correspond to $\pm10\%$ variation in the absolute value of $\left(\eta_{3\pi^0}^{-1}\right)$, while dotted lines correspond to a $\pm 10^{\circ}$
variation of its phase (with respect to
the expected value, i.e. $\left(\eta_{3\pi^0}^{-1}\right)\simeq\epsilon_S\simeq\epsilon$).
The value of $\left| \langle 3\pi^0 |T| \kln \rangle \right|^2$ has been kept fixed 
while varying $\left(\eta_{3\pi^0}^{-1}\right)$.}
   \label{fig:fig3}
\end{figure}
\begin{figure}[htbp] 
   \centering
   \includegraphics[width=5.5in]{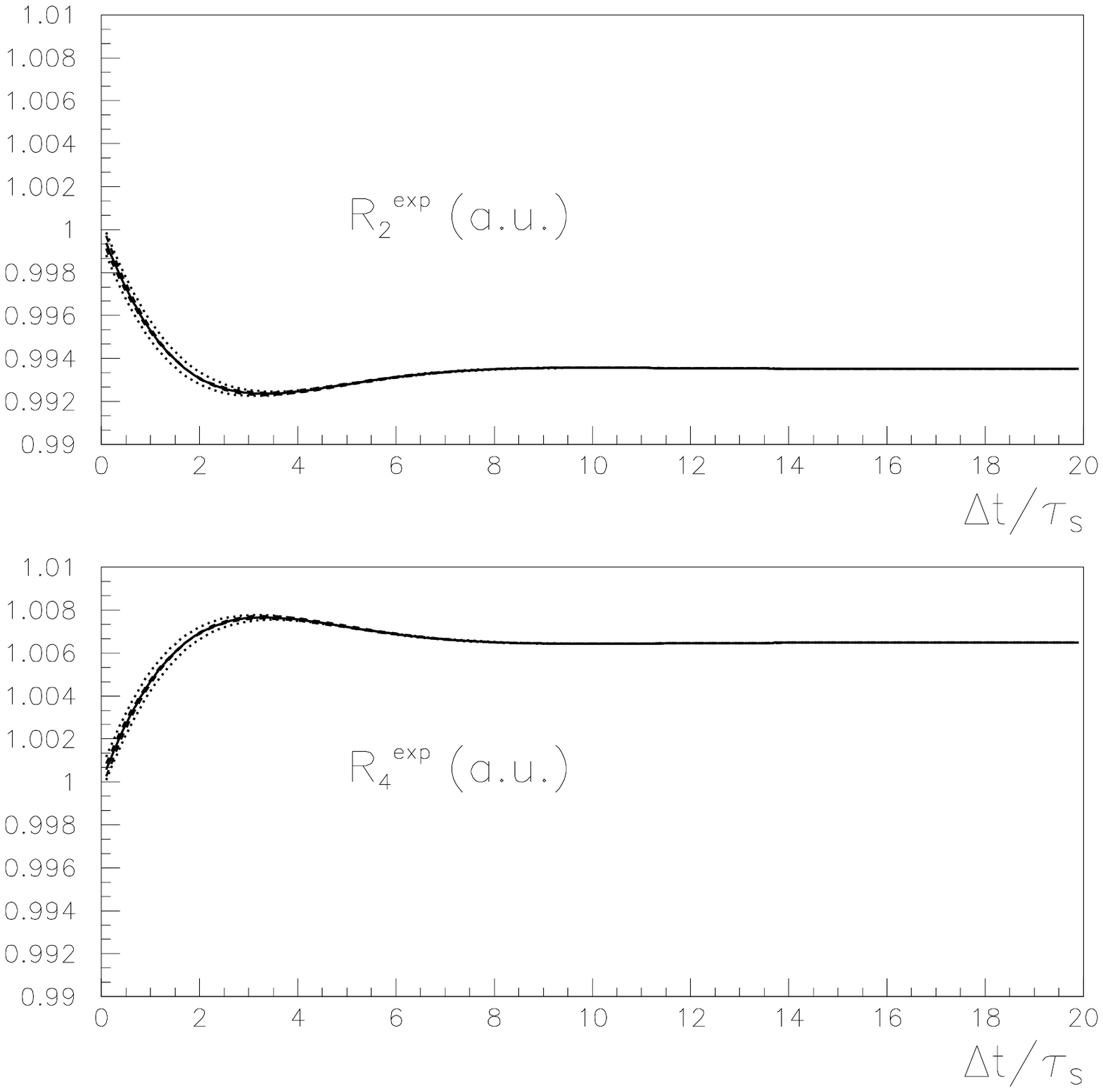} 
   \caption{A zoom of the plots shown in fig.\ref{fig:fig3} in the region $0\leq \Delta t \leq20 \tau_S$,
   which is statistically relevant for the KLOE-2 experiment at DA$\Phi$NE.}
   \label{fig:fig4}
\end{figure}

Even though it would be reasonable to expect also for the $\left(\eta_{3\pi^0}^{-1}\right)$
parameter a negligible contribution from direct \CP 
and \CPT violations \cite{ref:maiani,ref:wolf3pi}, i.e.
$\left(\eta_{3\pi^0}^{-1}\right)\simeq \epsilon_S$,
unfortunately the experimental knowledge on this parameter is much less precise than for
$\eta_{\pi\pi}$, resulting at present in an upper limit  $\left(\eta_{3\pi^0}^{-1}\right)< 9 \times 10^{-3}$ at 90\% C.L. \cite{ref:KLOE3pi0}.
However, also assuming 
a
contribution from direct \CP violation much larger than in the case
of $\pi\pi$, e.g. giving rise to a $\pm10\%$ variation in the absolute value of $\left(\eta_{3\pi^0}^{-1}\right)$, or a $\pm 10^{\circ}$
variation of its phase (with respect to
the expected value, i.e. $\left(\eta_{3\pi^0}^{-1}\right)\simeq\epsilon_S\simeq\epsilon$), the impact 
of these variations on the measured 
ratios $R_i^{\rm{exp}}(\Delta t)$ does not spoil the significance of the \T symmetry test in the 
$\Delta t$ region statistically relevant 
for the KLOE-2 experiment
at DA$\Phi$NE, 
i.e. 
$\Delta t \gtrsim-5 \tau_S$, 
as shown in Figs. \ref{fig:fig3} and \ref{fig:fig4}, 
where 
$\left| \langle 3\pi^0 |T| \kln \rangle \right|^2$ has been kept fixed 
to its measured value~\cite{ref:pdg2010}
while varying $\left(\eta_{3\pi^0}^{-1}\right)$.
%
Thus one can conclude that  direct \CP violation can be safely neglected. 
\par
Apart from these considerations, it is also possible to experimentally perform a direct test of assumption 
(\ref{eq:equiv}) 
by measuring the ratio of processes $\kn\to\kpp$ vs. $\kppp\to\knb$.
In fact, taking into account the difference between the tagged state $\kppp$ and the decaying state $\kpp$, using eq.(\ref{eq:probability}) one can easily evaluate the following ratio:
\begin{equation}
\dfrac{P[\kn(0)\to\kpp(\Delta t)]}{P[\kppp(0)\to\knb(\Delta t)]}\simeq
\dfrac{\left|e^{-i\lambda_S\Delta t}\left(\dfrac{1-\epsilon_L}{\sqrt{2}}\right)+e^{-i\lambda_L\Delta t}
{(\eta_{\pi\pi})}
\left(\dfrac{1-\epsilon_S}{\sqrt{2}}\right)\right|^2}
{\left|e^{-i\lambda_S\Delta t}\left(\dfrac{1-\epsilon_S}{\sqrt{2}}\right)+e^{-i\lambda_L\Delta t}
(\eta_{3\pi^0}^{-1})
\left(\dfrac{1-\epsilon_L}{\sqrt{2}}\right)\right|^2}~,
\end{equation}
which is constrained to be 1 if the condition $\eta_{\pi\pi}=(\eta_{3\pi^0}^{-1})$ holds, 
with the assumption 
of \CPT invariance ($\epsilon_S=\epsilon_L=\epsilon$).
Thus measuring this ratio with enough precision, one can evaluate whether the direct \CP violation contribution to the $3\pi^0$ decay is negligible, or not.
Analogous considerations apply to other ratios like:
\begin{itemize}
\item $P[\knb(0)\to\kpp(\Delta t)]/P[\kppp(0)\to\kn(\Delta t)]$
\item $P[\kn(0)\to\knn(\Delta t)]/P[\knnp(0)\to\knb(\Delta t)]$
\item $P[\knb(0)\to\knn(\Delta t)]/P[\knnp(0)\to\kn(\Delta t)]$
\end{itemize}